\documentclass{ws-procs9x6}

\begin{document}

\newcommand{\OVII}{O\ {\footnotesize VII}}
\newcommand{\OVIII}{O\ {\footnotesize VIII}}

\title{The Cluster Soft Excess: new faces of an old enigma}

\author{R. Lieu \& J.~P.~D. Mittaz}

\address{Physics Department, UAH\\
Huntsville\\
AL 35899, USA\\
E-mail: lieur@cspar.uah.edu}

\maketitle

\abstracts{
Until the advent of XMM-Newton, the cluster soft excess (CSE) was the subject
of some controversy due to both data analysis issues and uncertainties with the
soft excess emission mechanism.  XMM-Newton observations have finally laid to
rest any doubts as to the existence of the CSE and have also given tantalising
clues as to the nature of its emission mechanism.  Here we report on the
analysis of XMM-Newton observations of a number of CSE clusters in an attempt
to improve the analysis and understanding of the CSE.  Included as part of the
study is an analysis of the effects of background subtraction, which calls to
question the integrity of the claimed O VII line discovery, though not the soft
excess itself.  We also give details of both thermal and non-thermal fits to
the CSE cluster Abell 3112.}

\section{Introduction}

The cluster soft excess is an excess of flux seen above the hot ICM in a number
of clusters first discovered by Lieu et al. (1996), but its origin still
remains a mystery.  Currently there are two prevalent models, one thermal,
normally assumed to arise from warm ($\sim 0.2$keV) gas, and one non-thermal.
Observations with XMM-Newton have the potential to provide vital new
information to resolve this issue and recently there have been claims of Oxygen
lines associated with the CSE (Kaastra et al. 2003, Finoguenov, Briel \& Henry
2003).  If true this is very important to our understanding of the CSE since it
implies that its origin must be thermal rather than non-thermal.  However,
there are several issues that first need to be addressed before we can be sure
that the emission is indeed thermal.  The most important is that of background
subtraction - to date most CSE studies have used the background dataset of Lumb
et al. (2002) exclusively and therefore have not taken into account Galactic
background variations across the sky.  As we will show, this can cause an
overestimate of the CSE as well as the spurious detection of \OVII.

\section{Using in-situ backgrounds}

To investigate if variations in the Galactic background are important to the
analysis of XMM-Newton data on the CSE we must determine in-situ backgrounds
i.e. we must use observations which contain uncontaminated background signal.
This will either be when the clusters fit inside the XMM-Newton field of view,
or offset pointing which contain both cluster and background flux, Of all the
CSE clusters currently studied with XMM-Newton to data, only four fit
approximately within the XMM-Newton field of view and we discuss two of these,
Abell S1101 and Abell 3112.  These two clusters are amongst the most important
CSE clusters: Abell S1101 has the largest reported ROSAT PSPC soft excess
(Bonamente et al. 2001) and Abell 3112 has one of the largest XMM-Newton soft
excesses (Nevalainen et al. 2003).  For both clusters the estimated
contamination of the background by residual cluster emission in the 12'-14'
annulus used to estimate the background is small.  For these two cluster we can
therefore use a double subtraction method (see for example Arnaud et al. 2002)
to remove the correct background.

\begin{figure}
\centerline{\includegraphics[height=2.3in,angle=-90]{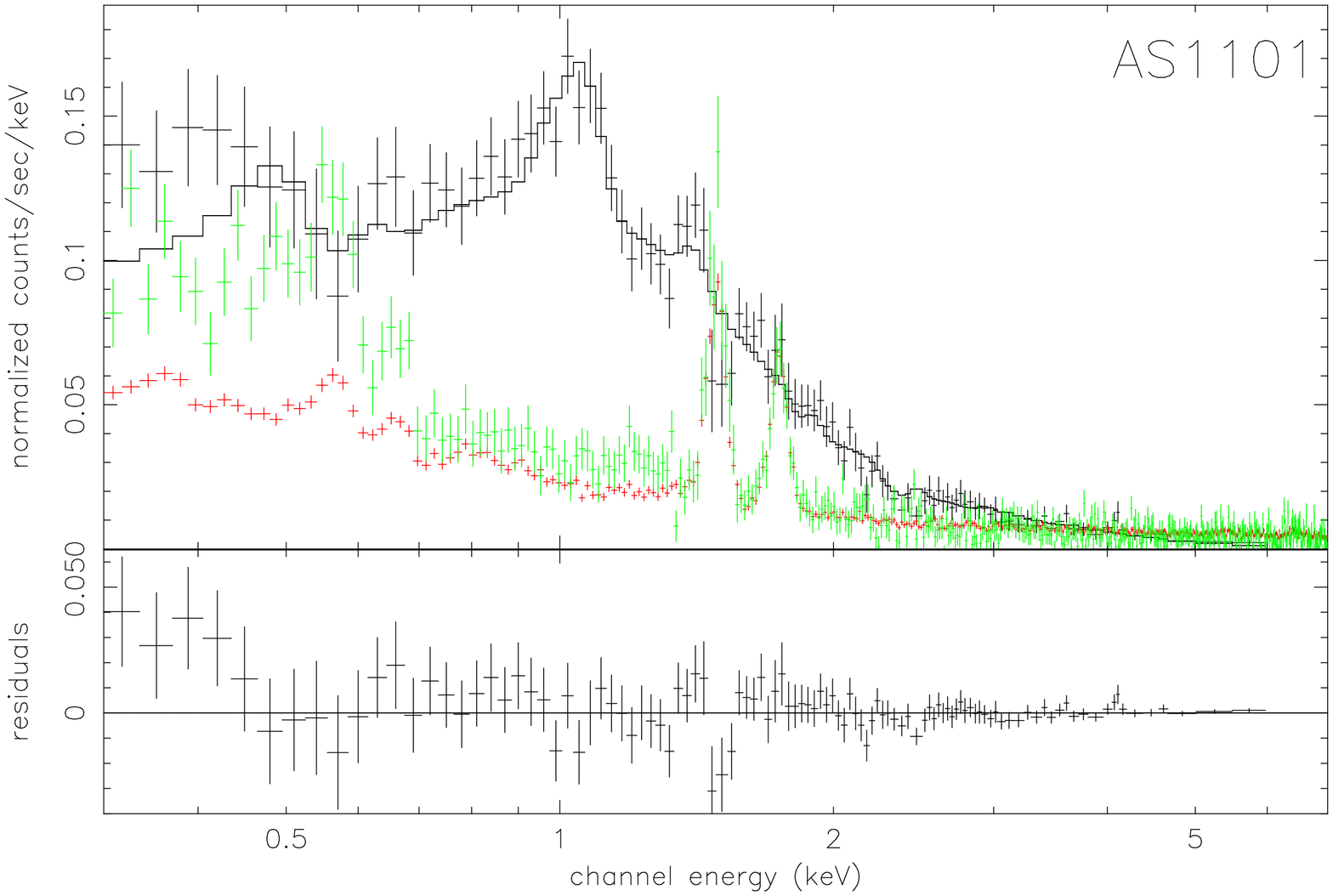}\hspace{0.2cm}\includegraphics[height=2.3in,angle=-90]{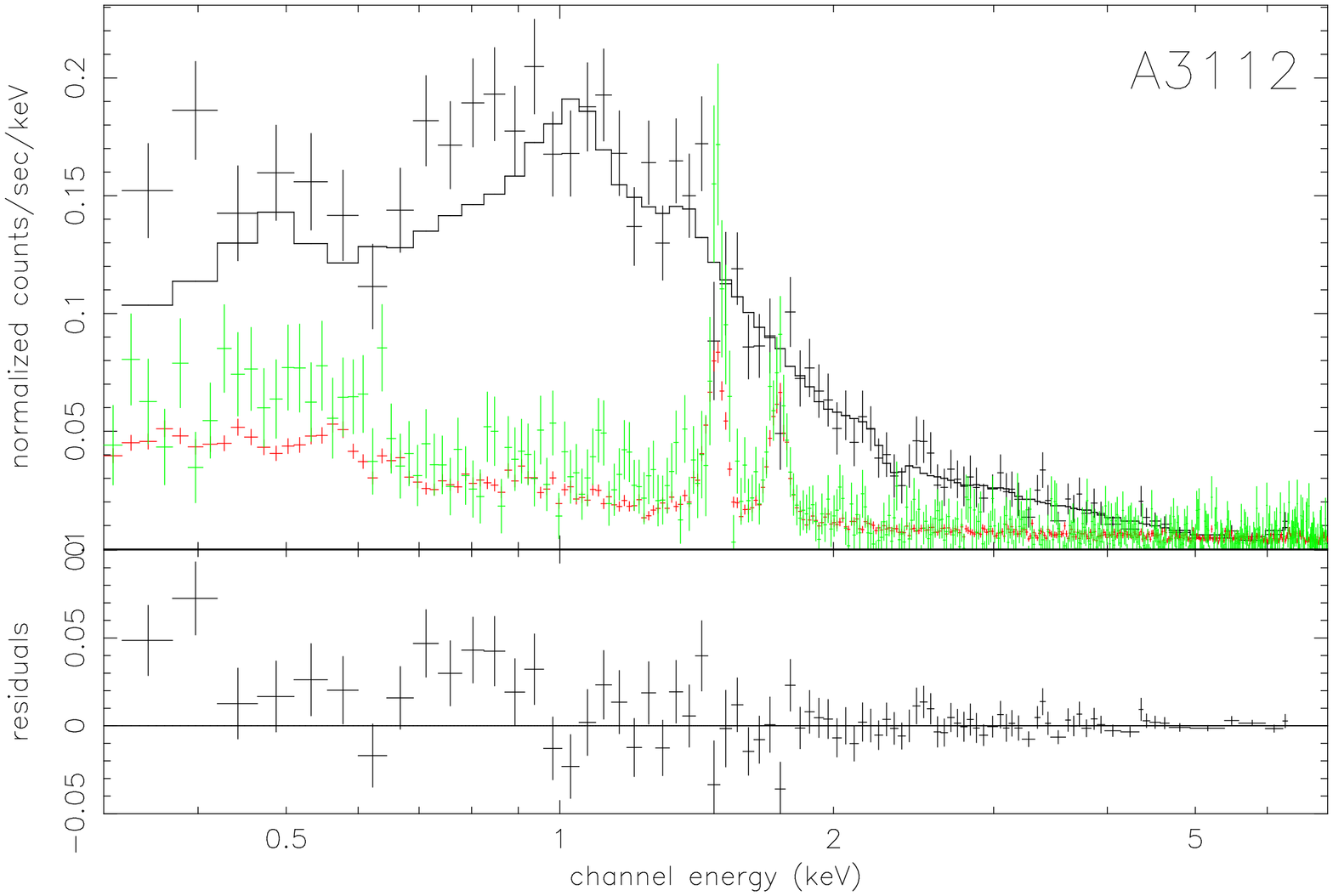}}
\caption{The top spectrum shows the background subtracted spectra from EPIC
  MOS1 for the 4'-6' arcminute annulus for AS1101 and A3112.  The other spectra
  show the Lumb background (lowest spectrum) and the double subtracted
  background (middle spectrum).  All backgrounds have been normalised to the
  10-12 keV source flux.  Note the strong \OVII line seen in the AS1101
  background spectra.  The solid line is a model for the hot ICM fitted to the
  1 - 7 keV band of the background subtracted spectrum.}\label{fig2}
\end{figure}

From the point of view of previous analyses of these clusters it is important
to compare the different background subtraction techniques.  We have therefore
compared the Lumb et al. (2002) background used in, for example, Kaastra et
al. (2003,2004) (lowest spectrum) with the double subtraction background method
using an in-situ background (middle spectrum).  Figure~\ref{fig2} shows that
there is considerable difference between the Lumb background and the double
subtraction background particularly in the case of Abell S1101.  Using the Lumb
et al. background alone can give rise to a spurious soft X-ray signal which
could be confused with a soft excess.  However, the model fit to the double
background subtracted spectrum (shown as a solid line on the top spectra in
Figure~\ref{fig2} together with the fit residuals show that there is still
evidence for a soft excess.  Therefore, even taking into account the correct
Galactic background subtraction, the soft excess still persists.

\section{The \OVII/\OVII\ line region}

One of the key findings of the Kaastra et al. (2003,2004) works was the
apparent detection of \OVII\ in the outer parts of a number of clusters
including Abell S1101.  The line detection was seen at the outer regions of
clusters and Kaastra et al. directly associated the emission with the CSE.
They also related this emission with large extended soft X-ray emission seen in
ROSAT all sky survey maps, directly linking the emission with the supercluster
environment and hence to the Warm Hot Intergalactic Medium (WHIM).  However, in
our analysis we find no such line, although the {\it background} spectrum of
the Abell S1101 field (12'-14') shows strong \OVII\ emission.  At first sight,
the detection of this line seems to verify the claims of Kaastra et al. but
this can be directly tested as the \OVII\ is strong enough to be resolved and
we can therefore determine the redshift of the line directly.

\begin{figure}
\includegraphics[height=1.6in]{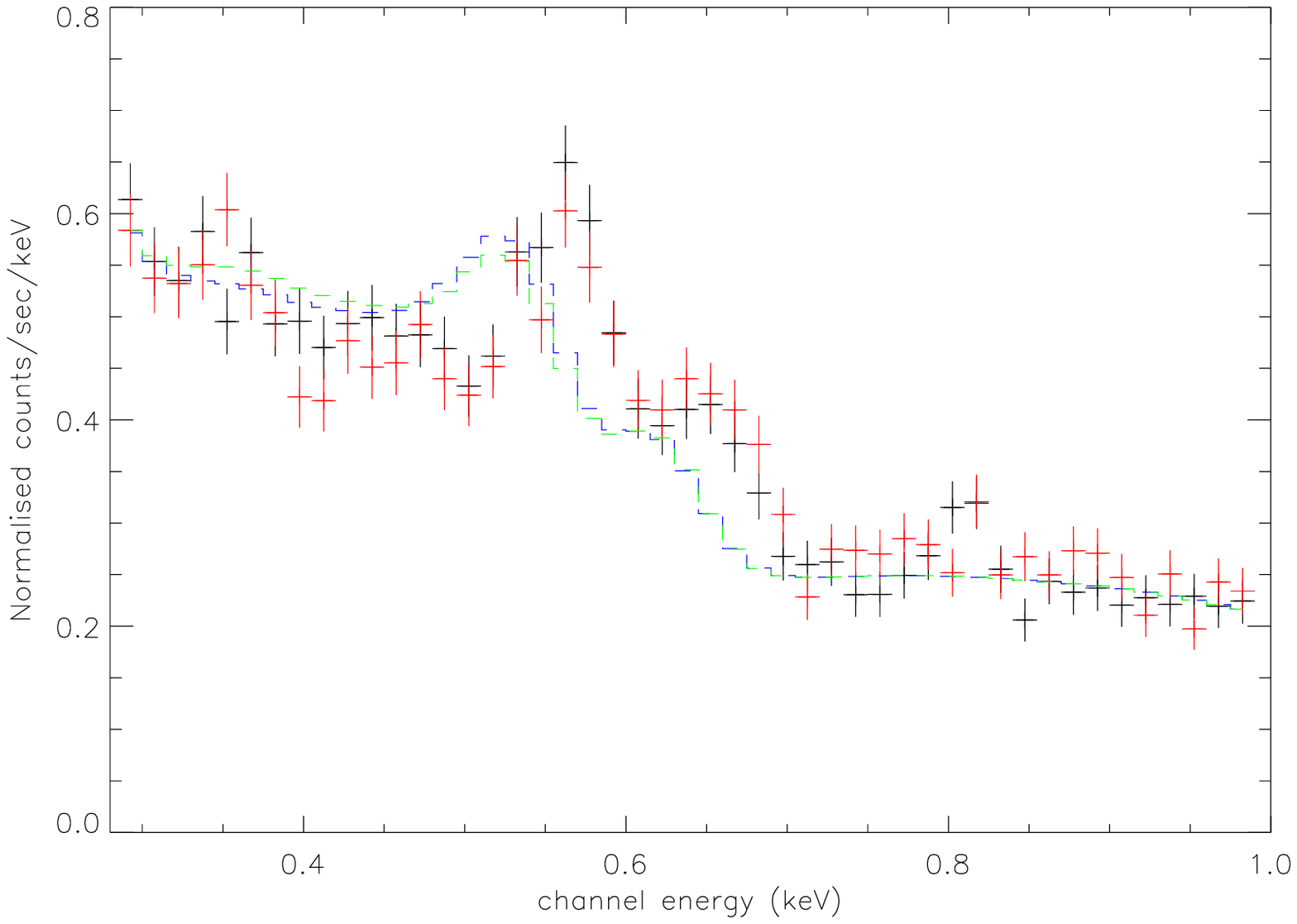}\includegraphics[height=1.6in,angle=0]{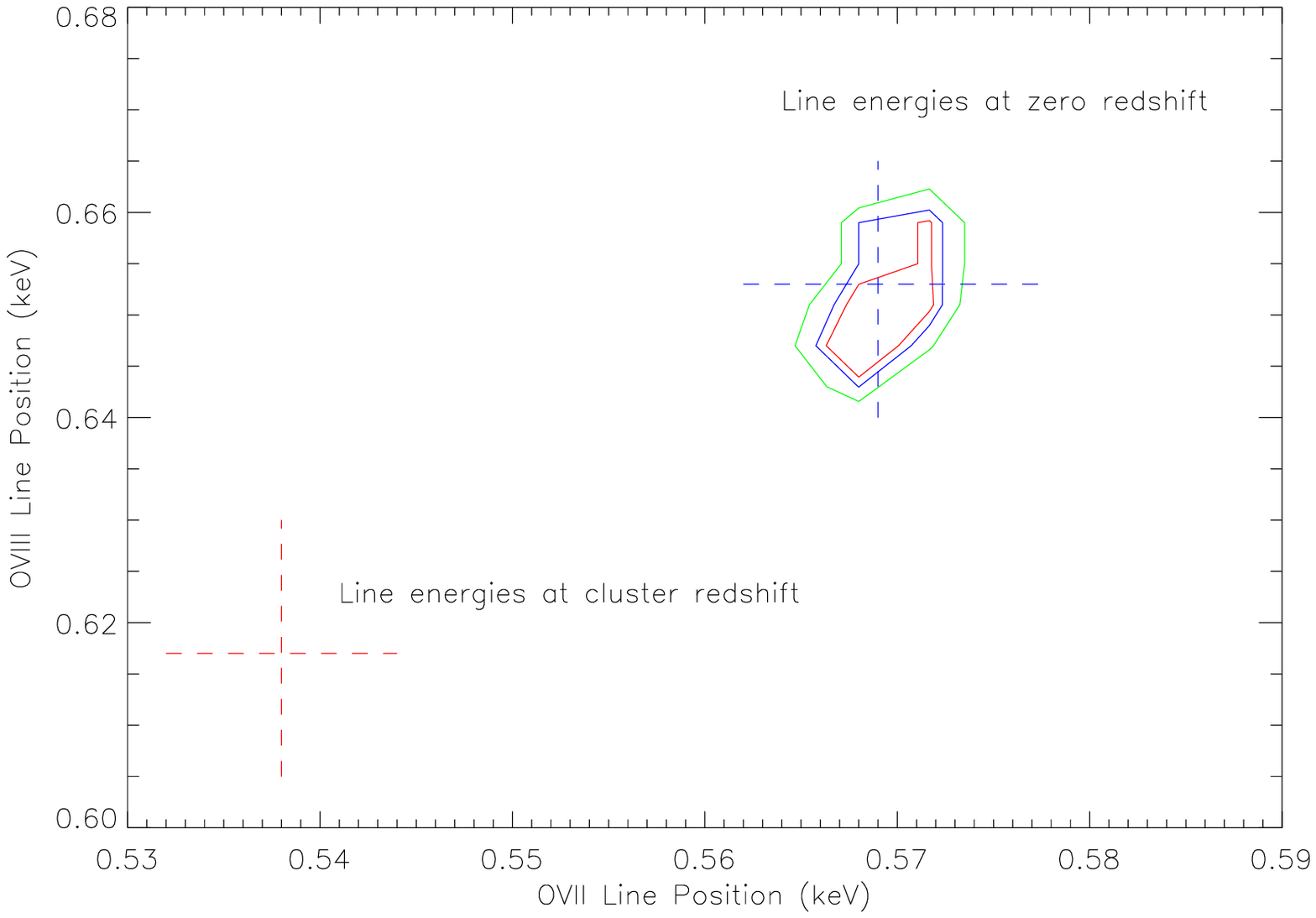}
\caption{The left panel shows the \OVII/\OVIII line region of Abell S1101
showing the two MOS instruments (black = MOS1, red = MOS2), together with
models including gaussians placed at the redshift of Abell S1101 (z = 0.058).
It is obvious that the redshifed Oxygen lines do not represent the observed
Oxygen lines.  The right panel shows the uncertainties to the energies of the
\OVII\ and \OVIII\ lines based on fits to the XMM-Newton data.  The lines are
clearly at the Galactic (zero) redshift rather than the redshift of the
cluster}\label{ovii-as1101}
\end{figure} 

The left panel of Figure~\ref{ovii-as1101} shows the \OVII/\OVIII\ line region
together with a model including a continuum model with two gaussians placed at
the redshifted energy of the \OVII/\OVIII\ lines.  It is clear that putting the
Oxygen lines at the redshift of the cluster does not match the observed
location of the lines.  The right hand panel of Figure~\ref{ovii-as1101} shows
the case where the lines have been put at zero (Galactic) redshift and it is
clear that the lines are much more consistent with a Galactic redshift rather
than the cluster redshift.  Indeed, if we fit the observed line energies we get
$0.569 \pm 0.003$ keV for \OVII\ and $0.651 \pm 0.007$ keV for \OVIII,
completely consistent with zero (Galactic) emission and completely inconsistent
with the redshifted line energies 0.538 keV and 0.618 keV.  We must therefore
conclude that the line emission and hence the majority of the emission at the
12'-14' annulus is Galactic in origin and therefore not associated with the
cluster.  This then not only verifies our use of the double subtraction method
since it shows that the background annulus is dominated by background emission,
but calls doubt on the claims of Kaastra et al. (2003,2004) of the association
of the CSE with an extended supercluster WHIM emission component.

\section{A case study: Abell 3112}

In order to investigate the impact of our background subtraction method on the
observed properties of the soft excess, we have first looked at the cluster
with the strongest XMM-Newton CSE, A3112.  This cluster was previously studied
by Nevalainen et al. (2003) using the Lumb data as the sole background
estimator.  Figure~\ref{fig2} shows that the Lumb background is not too
dissimilar to double subtraction background with the double subtraction
background being 30-40\% higher.

\begin{figure}
\centerline{\includegraphics[height=2.5in,angle=-90]{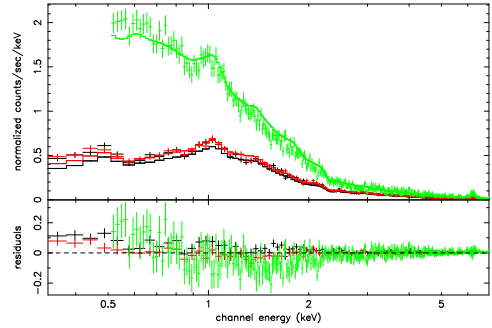}}
\caption{Single temperature broad band fit (0.3-7 keV) to the 1'-2' arcminute
  annulus for all three EPIC cameras for Abell 3112.  Clear low energy
  residuals and a very poor fit ($\chi^2_\nu = 1.52$ null hypothesis $=
  10^{-16}$) show the presence of a strong soft excess.}\label{a3112-bb}
\end{figure}

\begin{figure}
\centerline{\includegraphics[height=2.3in,angle=-90]{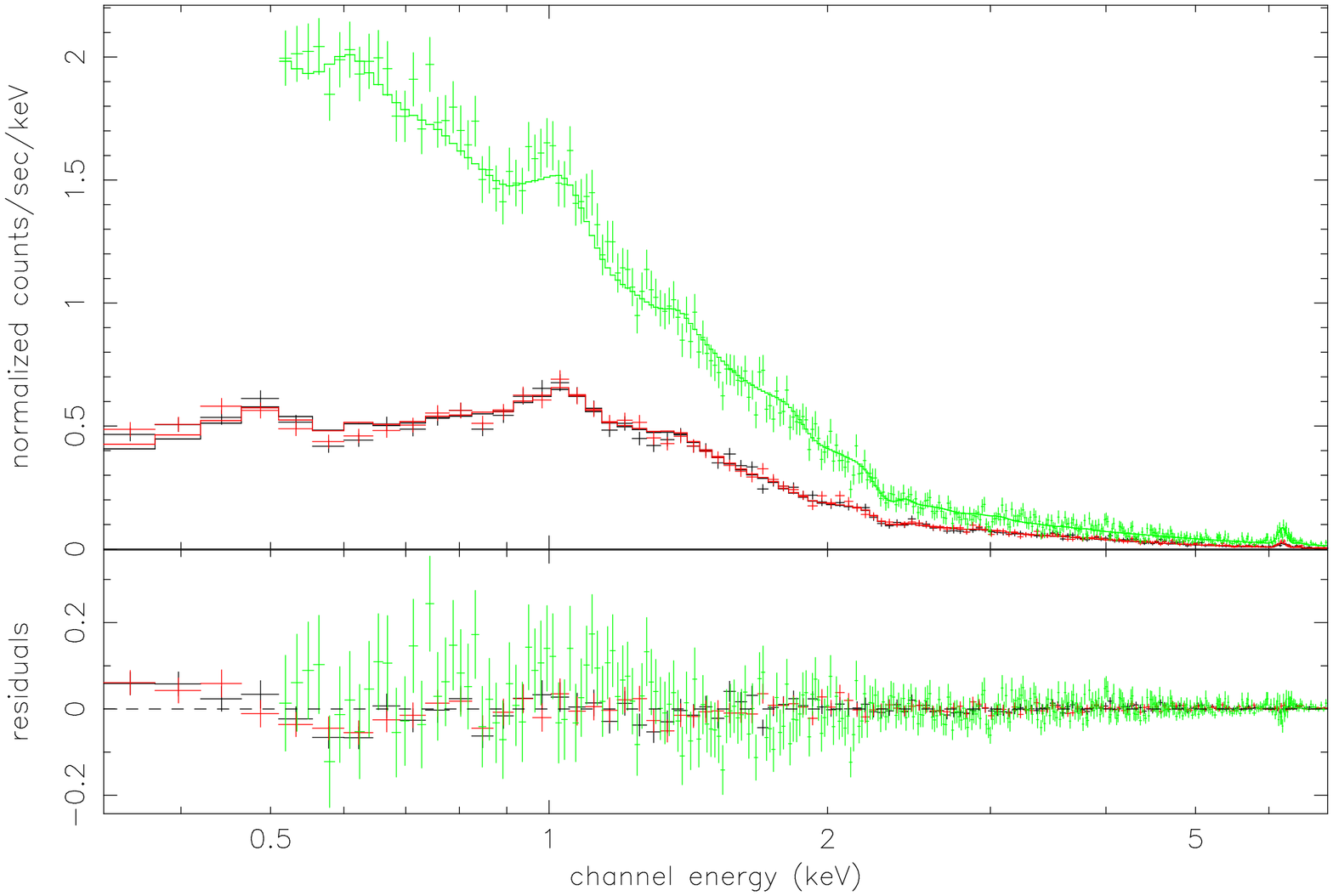}\hspace{0.2cm}\includegraphics[height=2.3in,angle=-90]{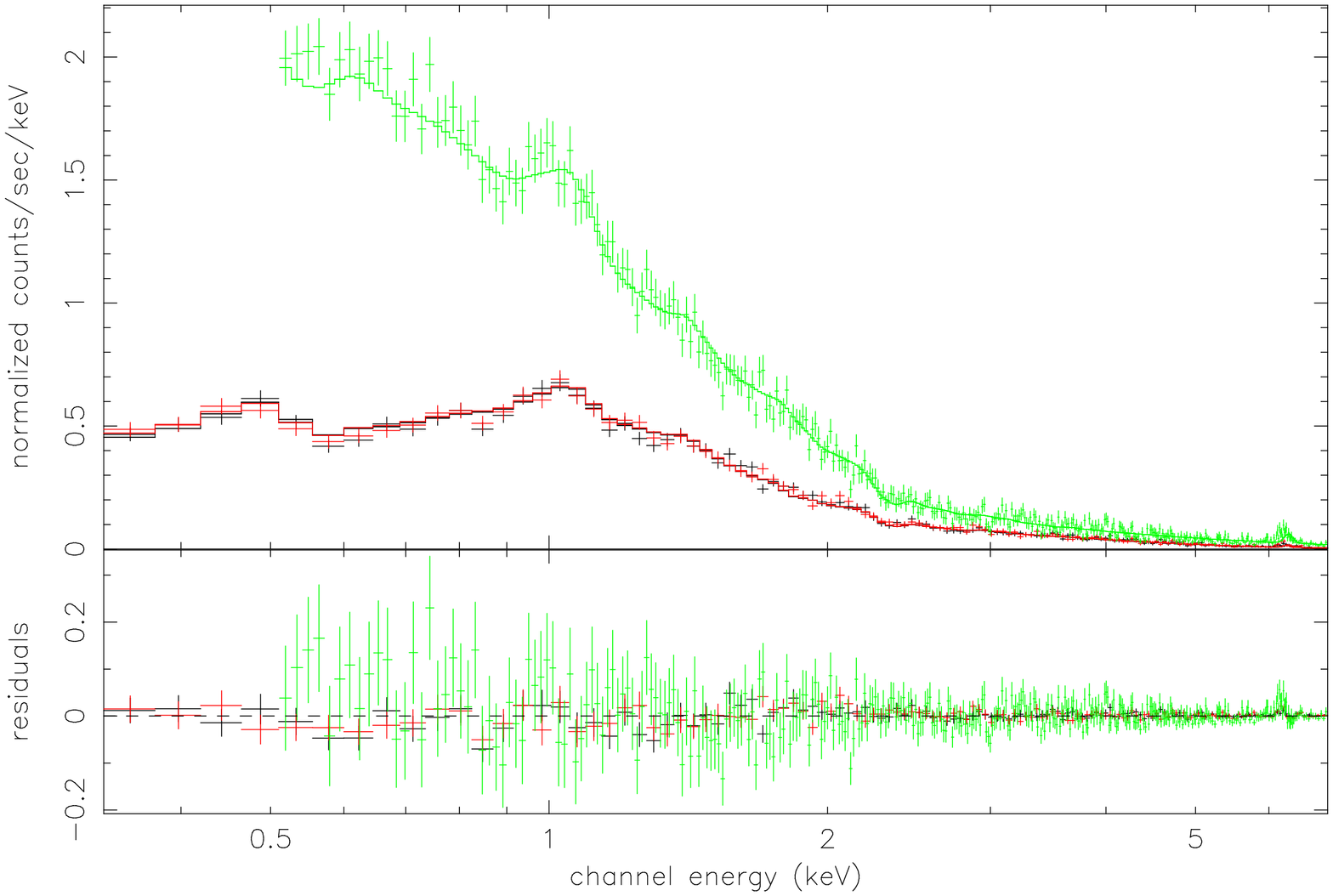}}
\caption{Left panel shows a two temperature model fit to the data (a thermal
soft excess) and the right panel shows a hot ICM plus power-law (non-thermal)
fit to the same data.  The $\chi^2$ values for these fits are $\chi^2_\nu =
1.03$ (null hypothesis = 0.28) for the thermal soft excess and $\chi^2_\nu =
1.05$ (null hypothesis = 0.16) for the non-thermal model.  Both models have
improved significantly over a single temperature fit.}\label{a3112-2mod}
\end{figure}

Figure~\ref{a3112-bb} shows a broad band (0.3-7 keV) single tempertature fit to
all three EPIC instruments to the 1'-2' arcminute annulus.  Note that we have
only fitted the 0.5 - 7 keV band for the PN due to calibration problems below
0.5 keV (Kirsch et al. 2004).  This single temperature fit is very poor with
strong residuals at low energies and a bad $\chi^2 = 1.52$.  These are all
characteristics of the presence of a strong soft excess in this cluster.  We
have therefore fitted extra models to the spectrum, on representing a thermal
soft excess, one representing a non-thermal model.  The fits are shown in
Figure~\ref{a3112-2mod} with the $\chi^2$ values for these fits are $\chi^2_\nu
= 1.03$ (null hypothesis = 0.28) for the thermal soft excess and $\chi^2_\nu =
1.05$ (null hypothesis = 0.16) for the non-thermal model respectively.  Both
models have improved significantly over a single temperature fit.  However, it
is not possible to distinguish between a thermal or non-thermal model.
Hopefully, a more detailed analysis of this and other CSE clusters with
XMM-Newton will be able to resolve the true emission mechanism.

\section{Discussion}

We have re-assessed the XMM-Newton data analysis of two clusters with reported
soft excesses.  We find that using a separate background dataset without taking
into account the variation of the soft X-ray Galactic background by using
in-situ background measurements can lead to an underestimate of the true
background and therefore an overestimate of the cluster soft excess.  It can
also lead to the spurious detection of Galactic \OVII\ lines which in the past
have been ascribed to the clusters themselves.  This calls into question many
of the conclusions reported in Kaastra et al. (2003,2004).  We do find,
however, that with the correct background subtraction the cluster soft excesses
still exists even though no \OVII\ lines can be seen.  Fundamentally, this
demonstrates that if we are to correctly understand and parameterise the CSE
using XMM-Newton data, the correct background subtraction is absolutely vital.
By using the correct background subtraction we then find that instead of having
a clear cut emission model as has been recently proposed, we are back to the
situation where it is not clear what mechanism gives rise to the CSE.

We have also assessed the physicallity of the different models.  With regard to
the status of thermal emission from the CSE we cannot completely discount it,
but there is no definitive evidence supporting it.  On the non-thermal side,
emission at the centre may require a non-thermal explanation, but on the
outskirts it is not clear if there can be enough energy inputted into the
relativistic CR population to give rise to the strength of the CSE.  Hopefully,
new observations and new instruments such as Astro-E2 will be able to finally
resolve this issue.


\begin{thebibliography}{99}

\bibitem{Arnaud} Arnaud, M., Neumann, D., Aghanim, N., Gastaud, R.,
Majerowicz, S., Hughes, J., 2001, A\&A, 365, L80
\bibitem{Bonamente1101} Bonamente, M., Lieu, R. \& Mittaz, J., 2001, ApJ,
561, L63
\bibitem{Kaastra03} Kaastra, J.S., Lieu, R., Tamura, T., Paerels, F.B.S., den
Herder, J.W., 2003, A \& A, 397, 445
\bibitem{Kaastra04} Kaastra, J.S., Lieu, R., Tamura, T., Paerels, F.B.S.,
den Herder, J.W., 2004, in `Soft X-ray emission from clusters of
galaxies and related phenomena', eds. R. Lieu \& J, Mittaz (Kluwer Dordrecht), 
p37
\bibitem{Kirsch} Kirsch, M.G.F., Altieri, B.,  Chen, B., Haberl, F., Metcalfe,
L., Pollock, A.M.T., Read, A.M., Saxton, R.D., Sembay, S., Smith, M.J.S.,
2004, astro-ph/0407257
\bibitem{Lieu96} Lieu, R., Mittaz, J.P.D. et al., 1996, Ap,J.\rm, 458, L5
\bibitem{Lumb} Lumb, D.H., Warwick, R.S., Page, M., DeLuca, A., 2002, A\&A,
389, 93
\bibitem{Nevalainen} Nevalainen, J., Lieu, R., Bonamente, M., \& Lumb, D.
2003, ApJ, 584, 716

\end{thebibliography}
\end{document}